\documentstyle[editedvolume,epsfig,namedreferences]{crckapb} 

\begin{opening}
\title{A comparison of radio emission from neutron star and black hole X-ray
binaries}

\author{R.~P.~Fender}
\institute{Astronomical Institute `Anton Pannekoek' \\
and Center for High Energy Astrophysics,\\
University of Amsterdam, Kruislaan 403,\\
1098 SJ Amsterdam, The Netherlands}

\end{opening}

\runningtitle{Radio emission from X-ray binaries}

\begin{document}


\section{Introduction : radio emission from X-ray binaries}

Radio emission has now been detected from $\sim 20$\% of all X-ray
binaries. In nearly all cases it has been found to be variable and
to display a nonthermal\footnote{Here `nonthermal' is taken to mean
arising from a non-Maxwellian particle distribution which
cannot be described by a single temperature}
spectrum, and synchrotron emission has been
established as the most likely emission mechanism. 

Radio outbursts from X-ray binaries generally follow a pattern of fast
rise and power law and/or exponential decay, with an evolution of the
spectral index ($\alpha = \Delta \log S_{\nu} / \Delta \log \nu$) from
`inverted' ($\alpha \geq 0$), probably arising in (partially)
optically thick emission, to optically thin ($\alpha \leq -0.5$), with
a corresponding shift in the peak of emission to lower frequencies.
This is in qualitative agreement with models of an expanding,
synchrotron-emitting cloud, as proposed by van der Laan (1966) for
outbursts of active galactic nuclei. Furthermore, in recent years the
time evolution of several outbursts have been imaged at high angular
resolution with arrays such as VLA, MERLIN and VLBA, and these images
reveal the outflow of radio-emitting matter along more
or less collimated paths at relativistic velocities. More recently
relatively stable radio emission from at least one persistent X-ray
binary has also been resolved into a jet-like structure.  As a result
it has become widely, if not universally, accepted that radio emission
from X-ray binaries arises in synchrotron-emitting jets. For detailed
reviews and references, see e.g. Hjellming \& Han (1995), Mirabel \&
Rodr\'\i guez (2000), Fender (2000a).

In this paper I compare, briefly, the properties of radio emission
from both transient and persistent neutron-star and
black-hole(-candidate) X-ray binaries.

\section{Transients}

\begin{figure}
\begin{center}
\leavevmode{\epsfig{file=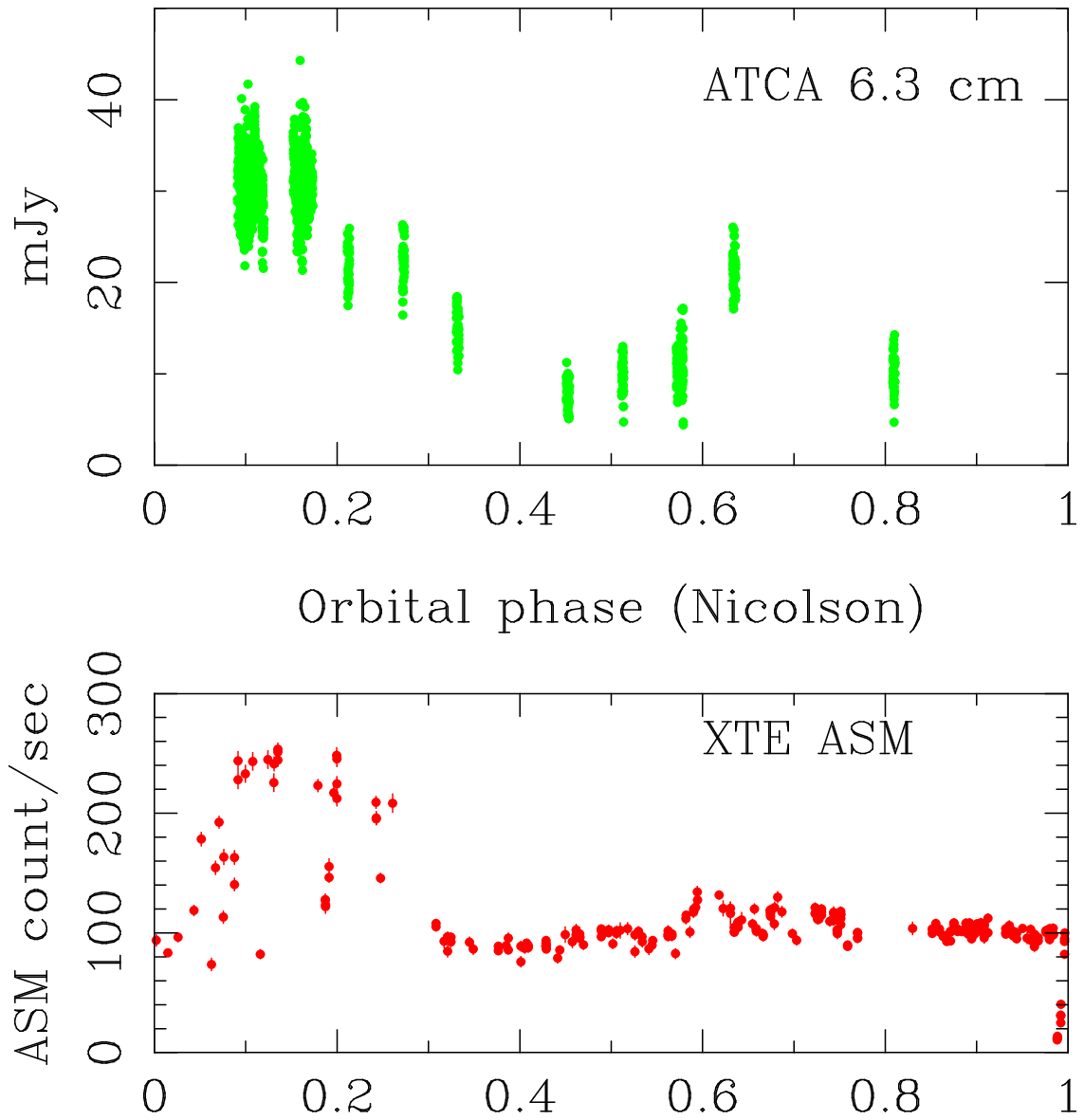,angle=0,width=6cm,clip=}\quad{\epsfig{file=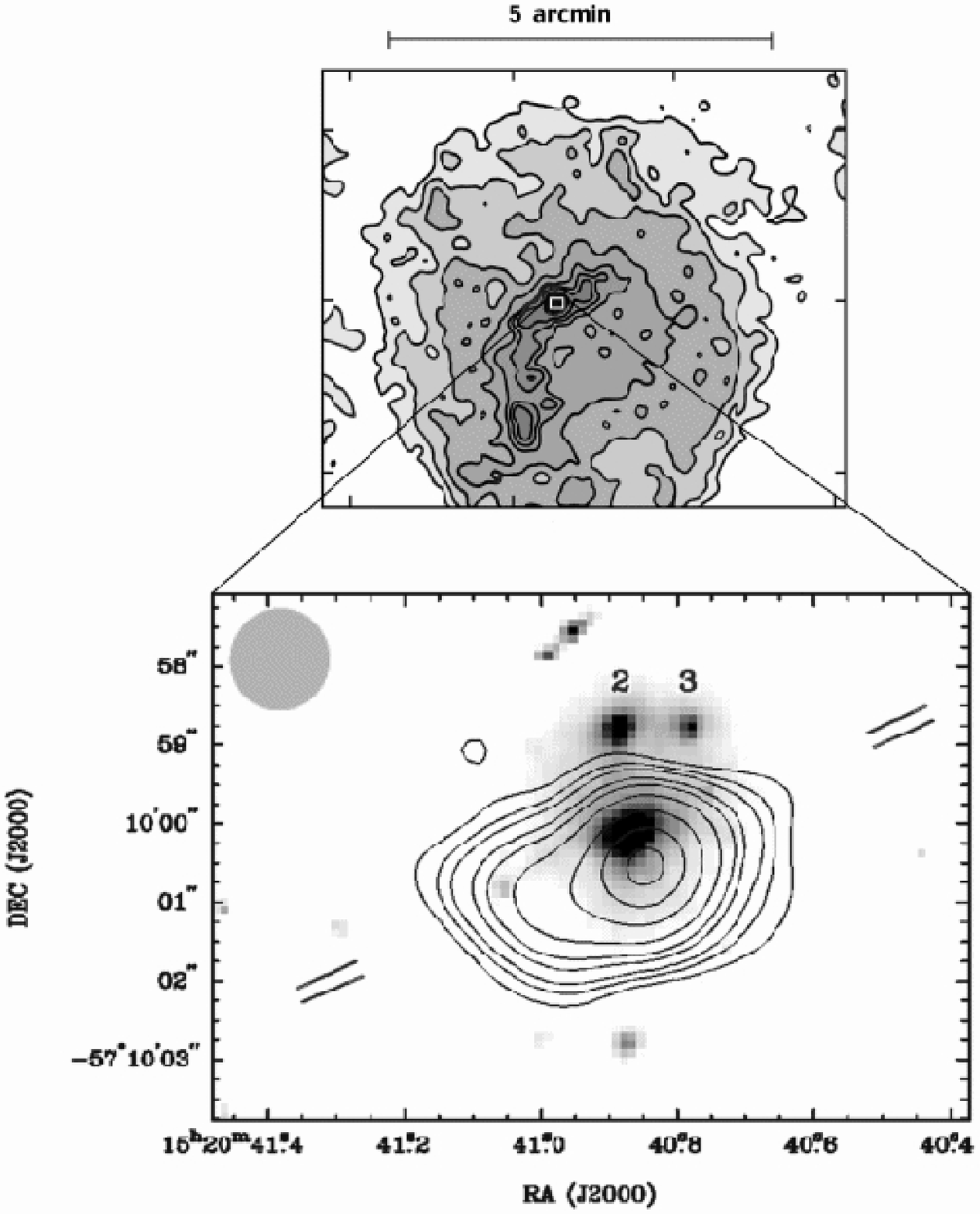,width=6cm,clip=}}}
\caption{Radio outbursts from the neutron star X-ray binary Cir
X-1. Left panel shows XTE ASM and radio monitoring around one 16.6-day
orbit; flaring presumably occurring during periastron passage of the
neutron star in an elliptical orbit (from Fender 1997). Right panel
shows arcmin- (top) and arcsec-scale radio jets (bottom) from the system (from
Fender et al. 1998).}
\end{center}
\end{figure}

Bright X-ray transients are generally accompanied by transient radio
emission which follows the pattern outlined in the
introduction.
Bright transients can contain both neutron stars
(e.g. Aql X-1) and black holes (e.g. GS 1124-684); do the radio
properties of these two populations differ?

The answer is yes and no. Firstly, it is clear that, excluding bright,
exotic objects for which classification of the compact object type has
proved impossible to date (e.g. Cyg X-3, SS 433, LS{\sc i}
+61$^{\circ}$ 303), the black hole transients are, at the peak of
outburst, the brightest radio sources associated with X-ray binaries.
However, this is also the case for their X-ray emission -- i.e. the
brightest transients in the X-ray band are also the black holes. It is
unclear at present whether this simply reflects the larger average
masses of the black holes, or differences in the accretion flows onto
the two types of compact accretor. On the other hand, the ratio of
radio to X-ray peak fluxes is comparable for both neutron star and
black hole X-ray binaries. As an example, the neutron star transient
Cen X-4 reached peak fluxes of $\sim 4$ Crab and $\sim 10$ mJy at soft
X-ray and radio wavelengths respectively during its 1979 outburst.
For comparison, the black hole transient A 0620-00 reached peak fluxes
of $\sim 45$ Crab and $\sim 200$ mJy during its 1975 outburst. While
the ratios are not exactly the same (but bear in mind there is likely
to be a large scatter in observed radio fluxes due to beaming -- see
Kuulkers et al. 1999), their order-of-magnitude correspondence
indicates that in both neutron star and black hole systems the ratio
of X-ray to radio luminosities is comparable. Furthermore, the ratio
is similar for most other systems (Fender \& Kuulkers, in prep).  This
in turn implies that the accretion and jet formation mechanisms are
broadly the same, during outburst, for both types of system.

\begin{figure}
\begin{center}
\leavevmode{\epsfig{file=all_rx.ps,width=8cm,clip=}\quad{\epsfig{file=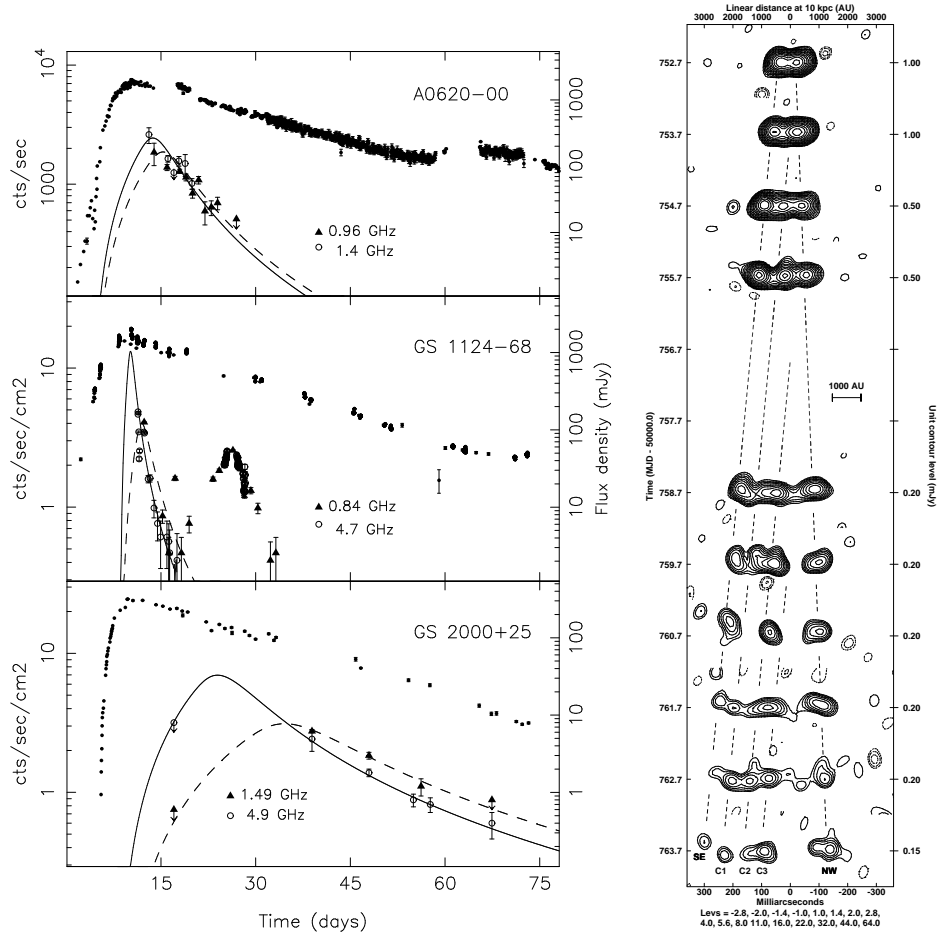,width=4cm,clip=}}}
\caption{Radio outbursts from black hole X-ray binaries. Left panel
compares X-ray and radio lightcurves of three black hole candidate
X-ray transients (from Kuulkers et al. 1999). Note the simularity of
the X-ray lightcurves (black dots) while radio light curves are
clearly different. Right panel illustrates
clearly resolved relativistic ejections from the black hole candidate
GRS 1915+105 (from Fender et al. 1999).}
\end{center}
\end{figure}

Cir X-1 stands out as an example of a neutron-star X-ray binary which
undergoes trasnsient radio-bright outbursts every 16.6 days (Fig 1).
The outbursts are believed to occur at periastron passage of a neutron
star in a highly eccentric orbit with a main sequence or slightly
evolved mass donor. The system displays a number of unique
characteristics including a radio jets which connect to a radio-bright
synchrotron nebula (Fender et al. 1998 and references therein),
rapidly evolving X-ray and radio light curves, and the highest
measured radial velocity ($\sim 400$ km s$^{-1}$) of any X-ray
binary (Johnston, Fender \& Wu 1999).
Given its relative brightness (core + jets are typically $\geq
10$ mJy) and the predictability of its radio outbursts, Cir X-1 may be
the best source in which to study the formation of jets by a
neutron-star accretor.

\section{Persistent sources}

Only four persistently bright X-ray sources in our Galaxy are believed
to contain black holes : Cyg X-1, GX 339-9, 1E 1740.7-2942 and GRS
1758-258. The latter two are too faint for regular monitoring, however
for Cyg X-1 and, especially, GX 339-4, we have a good idea of how
radio emission is related to X-ray state. This is explored in detail
in Fender (2000b), and is summarised in table 1. Discrete, bright,
outbursts correspond to major state transitions and/or the Very High
State; the formation of a steady outflow seems to occur in the
Low/Hard and, more weakly, Off states.

Neutron star systems do not display analogs of the black hole states,
but can nonetheless be classified into different groups. These are the
`Z' and `atoll' X-ray binaries (believed to contain low magnetic field
neutron stars), and the high-field X-ray pulsars. Thus the population
of neutron star X-ray binaries allows us to explore the effects of
mass accretion rate and accretor magnetic field on the production of
radio jets.  The Z sources (and the unusual atoll source GX 13+1) are
all regularly detected as variable radio source.

Penninx et al. (1988), in observations of the Z-source GX 17+2
established a link between radio emission and location on the Z track
(corresponding to the soft X-ray colours, and hence presumably the
state of the accretion disc, at the time of observation -- see
e.g. van der Klis 1995), such that radio emission is strongest on the
`horizontal branch', weak on the `normal branch' and absent on the
`flaring branch'. This relation is in agreement with (most) subsequent
studies, and at face value appears to demonstrate an anti-correlation
between radio emission and accretion rate in these sources, but this
is almost certainly an oversimplification.

The atoll sources are in general not detected except during outbursts
(e.g. Aql X-1 is an atoll source and a transient, which had a radio
outburst) and/or at very high mass accretion rates (which may be the
cause in the case of GX 13+1). No X-ray pulsar has ever been detected
as a radio source.

\begin{table*}
\begin{tabular}{cc}
\hline
State & Radio properties \\
\hline
Very High & Bright ejections with spectral
evolution from absorbed $\rightarrow$ optically thin \\
High/Soft & Radio suppressed by factor $\geq 25$ \\
Intermediate & Weak ? \\
Low/Hard & Low level, steady, flat spectrum extending to at least
sub-mm \\
Off       & Weak; similar to Low/Hard but reduced by a factor $\geq
10$ \\
\hline
\end{tabular}
\caption{The relation of radio emission to black hole X-ray
state. From Fender (2000b).}
\end{table*}

\begin{table*}
\centering
\begin{tabular}{ccccc}
\hline
Source & $S_{\nu} / (\rm{kpc}^2)$ & \multicolumn{3}{c}{Inferred
physical characteristics}\\
type   & (mJy) & ($\dot{m}/\dot{m}_{\rm
Edd})$ & B (Gauss)
& inner disc radius (km) \\
\hline
BHC (low/hard state) & $55 \pm 13$ & $\leq 0.1$ & -- & few $\times 100$ \\
Z (horizontal branch) & '' & 0.1 -- 1.0 & $10^9$--$10^{10}$
& few $\times 10$ \\
\hline
Atoll & $\leq 10 \pm 2$ & 0.01 -- 0.1 & $10^9$--$10^{10}$ & few $\times$ 10
 \\
X-ray pulsar & $\leq 6 \pm 2$ & $\leq 1.0$ & $\geq 10^{12}$ & $\geq 1000$ \\
\hline
\end{tabular}
\caption{
Comparison of derived mean intrinsic radio luminosities for the BHC/Z,
Atoll and X-ray pulsar classes of persistent X-ray binary, plus simple
interpretations of their physical differences. From Fender \& Hendry (2000).}
\end{table*}

These results, and their inferences, are summarised in table 2
(adapted from Fender \& Hendry 2000). It seems clear that an inner
($\leq 1000$ km) accretion disc, as a result of a low ($\leq 10^{10}$
G) accretor magnetic field and, most importantly and intuitively, a
high mass accretion rate, are required for jet formation.  In addition,
it is demonstrated in table 1 that the Z sources (plus GX 13+1) have
approximately the same radio luminosity as the persistent black hole
candidates when in the Low/Hard X-ray state. Once again we are forced
to conclude that the accretion and jet formation processes in neutron
star and black hole X-ray binaries are similar (although it is noted
that the Z sources are apparently a little more luminous in X-rays on
average than the black holes in the Low/Hard state).

\section{Conclusions}

This comparison of the radio properties of the neutron-star and
black-hole X-ray binaries has revealed that whether the systems are
{\em transient}, as a result (in most cases) of low average accretion rates
and disc instability mechanisms, or {\em persistent}, as a result of
high average accretion rates, the coupling between radio jet formation
and accretion luminosity is similar for both classes of accretor.
It appears that the disc : jet coupling does not really care too much
about the nature of the accretor (as long as the magnetic field is not
strong enough to disrupt the accretion disc, as in the case of the
X-ray pulsars).

However, this is an oversimplification of the situation, and many
important questions remain. For example, it is the atoll sources,
amongst the neutron star X-ray binaries which appear to show the
strongest X-ray spectral evidence for Comptonising coronae, yet they are
weak radio sources ... while in the black holes the presence of the
corona is (nearly) always associated with observable radio
emission. Is it just a question of accretion rate, or is 
another factor allowing only one of the classes to form jets ? This and
other questions will only be resolved by future coordinated radio and
X-ray observations of neutron-star, as well as black-hole, X-ray binaries.


\end{document}